\documentclass{iau-JDSS}
\usepackage{graphicx,natbib}

\begin{document}

\title[Spirals and ETGs]{Parallel-sequencing of early-type\\ and spiral galaxies}

\author[Michele Cappellari]{Michele Cappellari}

\affiliation{Sub-department of Astrophysics, Department of Physics, University of Oxford\\ Denys Wilkinson Building, Keble Road, Oxford OX1 3RH \\ email: {\tt cappellari@astro.ox.ac.uk}}

\pubyear{2012}
\jname{Highlights of Astronomy, Volume 16}
\editors{Thierry Montmerle, ed.}

\maketitle

\begin{abstract}
Since Edwin Hubble introduced his famous tuning fork diagram more than 70 years ago, spiral galaxies and early-type galaxies (ETGs) have been regarded as two distinct families. The spirals are characterized by the presence of disks of stars and gas in rapid rotation, while the early-types are gas poor and described as spheroidal systems, with less rotation and often non-axisymmetric shapes. The separation is physically relevant as it implies a distinct path of formation for the two classes of objects. I will give an overview of recent findings, from independent teams, that motivated a radical revision to Hubble's classic view of ETGs. These results imply a much closer link between spiral galaxies and ETGs than generally assumed.
\end{abstract}

\begin{figure}[h!]
\includegraphics[width=\textwidth]{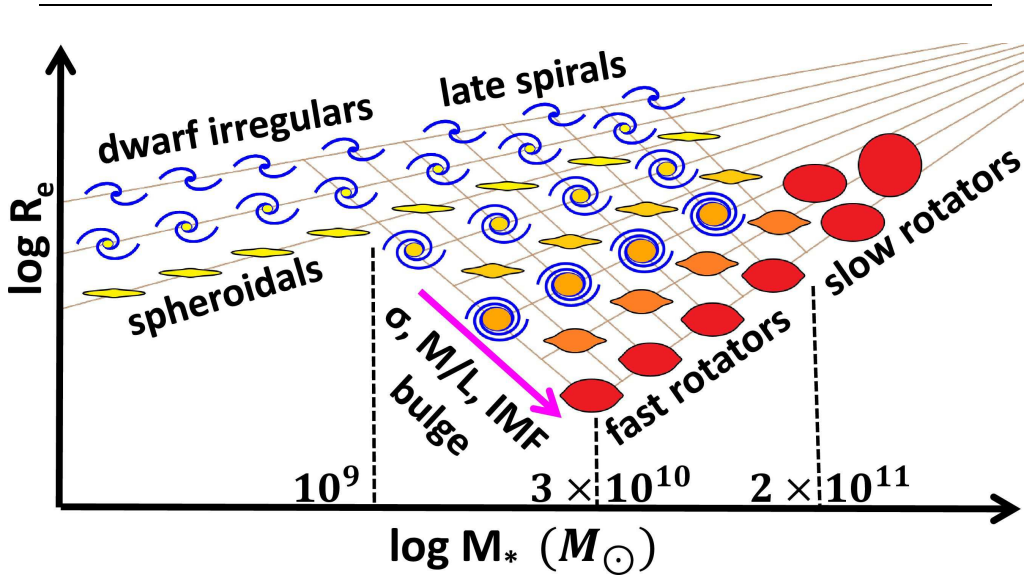}
  \caption{ETGs properties like shape, dynamics, population and IMF, merge smoothly with the properties of spiral galaxies on the mass-size diagram. All trends appear driven by an increase of the bulge fraction, which greatly enhance the likelihood for a galaxy to have his star formation quenched. This parallelism between the properties of spirals and ETGs motivated a proposed revision of Hubble's tuning-fork diagram. The same symbols are used in this figure (taken from \citealt{Cappellari2012p20}) as in the `comb' morphological classification diagram proposed in \citet{Cappellari2011b}.}
\end{figure}

\end{document}